\documentclass[prb,twocolumn,showpacs,preprintnumbers,amsmath,amssymb]{revtex4}
\usepackage{graphicx}
\usepackage{dcolumn} %for  two-column only
\begin{document}

\title{Step bunching of vicinal 6H-SiC\{0001\} surfaces}

\date{\today}

\author{Valery Borovikov}
\email{valery.borovikov@physics.gatech.edu}
\author{Andrew Zangwill}
\email{andrew.zangwill@physics.gatech.edu}
\affiliation{School of Physics \\ Georgia Institute of Technology,  Atlanta, Georgia 30332, USA}

\begin{abstract}
We use kinetic Monte Carlo simulations to understand  growth- and etching-induced step bunching of 6H-SiC\{0001\} vicinal surfaces oriented towards $<1 \overline{1} 00>$ and $<11 \overline{2} 0>$. By taking account of the different rates of surface diffusion on three inequivalent terraces, we reproduce the experimentally observed tendency for single bilayer height steps to bunch into half unit cell height steps. By taking account of the different mobilities of steps with different structures,  we reproduce the experimentally observed tendency for adjacent pairs of half unit cell height  steps  to bunch into full  unit cell height steps. A prediction of our simulations is that growth-induced and etching-induced step bunching lead to different surface terminations for the exposed terraces when full unit cell height steps are present.
\end{abstract}

\pacs{81.10.Aj, 81.15.Kk, 81.65.Cf}

\maketitle

\section {Introduction}

Silicon carbide (SiC) is a  very promising material for  microelectronic applications because of its superior electronic properties, high thermal
and chemical stability, high-power and high-frequency capability, and high tolerance to radiation damage.\cite{Matsunami, Fissel}
SiC is also an attractive candidate as a substrate for the heteroepitaxial growth of other materials.\cite{Davis, Okumura} A particulary exciting example (which motivated the present study) is the growth of epitaxial graphene by thermal decomposition of the basal surfaces of single crystal 4H and 6H SiC.\cite{Haas}  Nevertheless, SiC will not reach its anticipated potential until a variety of problems are solved, not least being the need to controllably grow device-quality single crystal material on a large scale.\cite{Matsunami, Fissel, Powell}

One approach to the growth problem is  ``step-controlled'' epitaxy, where new layers grown onto  surfaces vicinal to the hexagonal basal planes inherit the stacking order of the substrate through the step-flow mode of growth.\cite{Matsunami} Unfortunately, step-flow growth on vicinal surfaces does not always proceed by the uniform motion of a train of evenly spaced steps. Instead,  growth-induced step bunching often occurs, as it invariably does when vicinal surfaces are etched by exposure to hot hydrogen gas. Suggestions for the origin of SiC step bunching include impurity adsorption, \cite{Ohtani, Papaioannou}  differences in surface energetics for different bilayers of $ \alpha $-SiC polytypes, \cite{Heine, Chien, Kimoto, Tanaka} differences in intrinsic  step velocities and step configurations, \cite{Kimoto, Tanaka, Stout} and other differences in growth kinetics.\cite{Frisch, Heuell} However, no systematic exploration of any particular mechanism and comparison of the results with all available data seems to have been performed until now.

This paper reports the results of kinetic Monte Carlo (KMC) simulations  designed to identify the kinetic pathways that promote growth- and etching-induced step bunching of vicinal 6H-SiC surfaces.  We focus on surfaces vicinal to (0001) (Si-terminated) and ($000\overline{1}$) (C-terminated) with steps running perpendicular to the  $<1 \overline{1} 00>$ and $<11 \overline{2} 0>$ directions. These particular starting surfaces were chosen to make contact with experimental observations made using atomic force microscopy (AFM), low-energy electron diffraction (LEED) analysis, high-resolution transmission electron microscopy (HRTEM), and scanning tunnelling microscopy (STM).\cite{Kimoto, Nakamura, Nakamura2, Nakagawa, Nakajima, Feenstra, Hayashi} Our main conclusion is that the experimental results for the Si-face  are quite explicable  using a lattice model that recognizes that there are three inequivalent terraces for surface diffusion and two inequivalent steps with different mobilities.  The C-face data are similarly explicable (or at least rationalizable) if the terrace diffusion rates and step mobilities  are less different on this face than on the Si-face.

\section{Previous Work}
\subsection{Experimental Observations}
\begin{figure}
\begin{center}\includegraphics[
width=8.4cm, keepaspectratio] {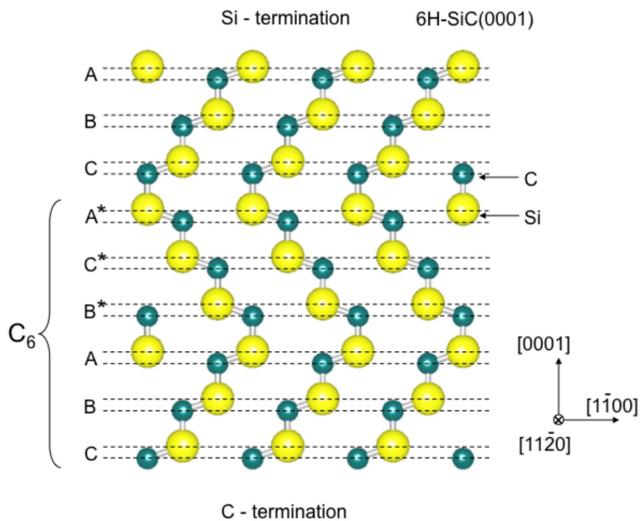}
\caption{\label{fig:6H-SiCatomicarrangement}  {Arrangement of Si and C atoms in 6H-SiC crystal. A fragment of $(11\overline{2}0)$ atomic plain is shown schematically.% $C_6$ indicates a single unit cell.
}}
\end{center}
\end{figure}

\begin{figure}
\begin{center}\includegraphics[
width=8.4cm, keepaspectratio] {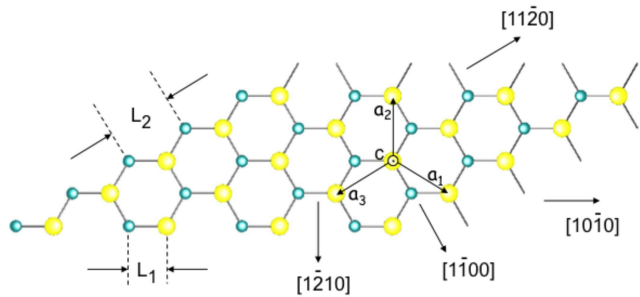}
\caption{\label{fig:directionsandunits}  {Top view of a single bilayer is shown schematically. $L_{1}$ and $L_{2}$ are the units of length in $<1\overline{1}00>$ and $<11\overline{2}0>$ directions, correspondingly.}}
\end{center}
\end{figure}

This section reviews the experimental observations of etching- and growth-induced bunching for surfaces vicinal to the basal planes of 6H-SiC. The designation 6H refers to the bulk unit cell indicated in Figure~\ref{fig:6H-SiCatomicarrangement} where six bilayers of silicon and carbons atoms are arranged in the particular stacking sequence shown.
If it were exposed to vacuum, the top layer of silicon atoms would be a typical $(0001)$ Si-terminated surface. The bottom layer of carbon atoms similarly exposed would be a typical $(000\overline{1})$ C-terminated surface.
Along the c-axis of 6H-SiC crystal, the bilayers are arranged in two groups of three bilayers each. Each bilayer is exactly the same apart from rigid lateral shifts within one group of bilayers and rotation of the lattice by $60^{\circ}$ around the c-axis from one group of bilayers (A, B, C) to the other ($A^{\ast}, C^{\ast}, B^{\ast}$). The hexagonal arrangement of atoms shown in Figure~\ref{fig:directionsandunits} is a (projected) view along $[0001]$ of one of the bilayers from the  $A^{\ast}C^{\ast}B^{\ast}$-group seen in edge-view in Figure~\ref{fig:6H-SiCatomicarrangement}.

There have been many experimental studies of growth\cite{Kimoto, Tanaka, Papaioannou, Ohtani, Syvajarvi, Kong, Ueda, Nakamura2} and etching\cite{Nakamura, Nakagawa, Nakajima, Feenstra, Hayashi} of 6H SiC surfaces cut slightly vicinal to its basal planes. The details vary (temperature, doping, and Si/C ratio during growth), but most papers focus on the Si-terminated face with miscuts oriented along the $<11 \overline{2} 0>$ direction. Only a few report results for the  C-terminated surface or for vicinal miscuts oriented along the $<1 \overline{1} 00>$ direction. For our purposes the systematic  H-etching experiments conducted by Feenstra and co-workers are  particularly valuable.\cite{Feenstra} Table~I  summarizes their AFM and HRTEM observations for nominally on-axis samples and samples with intentional miscuts of $\sim 3^\circ$ and $12^\circ$. The numbers ${1 \over 2}$ and $1$ in the table refer to observations of more-or-less ordered arrays of bunches where 3 Si-C bilayer steps have bunched into a single step with the height of ${1\over 2}$ unit cell and where two such bunches have further bunched into one unit cell height steps. Etching  of $(0001)$ surfaces miscut  $12^\circ$ along $<1\overline{1}00>$ produces increased bunching into $4$-$5$ unit cell height steps which may more properly be regarded as  ``nanofacets''.\cite{Nakagawa} For the same surface miscut $12^\circ$ along $<11\overline{2}0>$, the authors find no average step orientation due to large-scale step meandering. From the corresponding image of the C-face, we infer (hence the quotation marks in the table) similar, but less pronounced, step meandering.  Quite generally, the data summarized Table~I demonstrate that the tendency for etching-induced bunching is greater on the Si-face than  on the C-face and greater for steps oriented perpendicular to $<1\overline{1}00>$ than for steps oriented perpendicular to $<11\overline{2}0>$.
 \begin{table}
\begin{tabular}{lccc}
  \hline  \\
  & 6H SiC(0001) & {\rm Etching Results} & (Ref.~\onlinecite{Feenstra})\\ \\
  \hline \\
vicinal angle & $~<0.3^\circ$ & $~~~~ \sim 3^\circ$ &~~~~ $12^\circ$ \\ \\
 ~~~~~~~ & ~ &Si-terminated & ~ \\ \\
  $[1 \overline{1} 00]$~~~~~~~ &~ 1 &~~~~ 1 &~~~~ 4-5 \\ \\
  $[11 \overline{2} 0]$~~~~~~~ &~ ``1/2" &~~~~ 1/2 &~~~~ MEANDER \\ \\
 ~ & ~ &C-terminated  & ~ \\ \\
  $[1 \overline{1} 00]$~~~~~~~ &~ 1/2 &~~~~ ``1" &~~~~ 1 \\ \\
  $[11 \overline{2} 0]$~~~~~~~ &~ ``1/2" &~~~~ 1 &~~~~ ``meander" \\
  \hline

\end{tabular}
\caption{ {Experimentally observed etching-induced step bunching of $(0001)$ (Si-terminated) and $(000\overline{1})$ (C-terminated) surfaces of 6H-SiC for various miscut angles and orientations. Entries in quotation marks are inferred by the present authors. See text for discussion.}}
\end{table}

There is no single data set for growth-induced step bunching comparable to the etching-induced results summarized in Table~I. Nevertheless, a survey of the literature reveals trends very similar to the etching data. Thus, for surfaces miscut by $3.5^\circ$ toward $<11\overline{2}0>$,  Kimoto {\it et al.}\cite{Kimoto} find that growth on Si-terminated surfaces produces  ${1\over 2}$ unit cell height bunches while growth on comparable C-terminated surfaces is twice as likely to remain completely unbunched (only SiC bilayer steps appear) as to bunch into ${1\over 2}$ unit cell height steps. Similarly, data obtained for step-flow growth on vicinal Si-terminated surfaces oriented toward $<1\overline{1}00>$ exhibit 6 bilayer bunches (full unit cell), compared to only 3 bilayer bunches (half unit cell) observed for similar surfaces miscut along $<11\overline{2}0>$.\cite{Nakamura2} Therefore, as during etching, the tendency for growth-induced bunching is greater on the Si-face than  on the C-face and greater for steps oriented perpendicular to $<1\overline{1}00>$ than for steps oriented perpendicular to $<11\overline{2}0>$. For both growth and etching, step bunching is always more pronounced on surfaces with higher miscut angles.

\subsection{KMC Simulations}
Three groups have used kinetic Monte Carlo simulations to study step flow growth on vicinal SiC\{0001\} surfaces. Heuell used a one-dimensional model that did not distinguish carbon atoms from silicon atoms.\cite{Heuell} Two inequivalent types of steps were considered and the probabilities for a diffusing adatom to attach to each step from the terraces below and above were treated as independent parameters. A parameter set was found were an initial train of height-one steps bunched into a train of height-six steps.
However, trains of ${1\over 2}$ unit cell height steps are commonly observed in experiments, which suggests that a train of single bilayer steps bunches first into a train of ${1\over 2}$ unit cell height steps, which then bunch  into full unit cell height steps.\cite{Kimoto, Feenstra} Heuell's model does not produce this behavior.

Stout developed a very elaborate KMC simulation that took account of the  SiC crystal structure, the transport, adsorption, and surface diffusion of physisorbed precursors, the dissociative chemisorption, surface diffusion, and desorption of  dissociated species, and the attachment/detachment of adatoms to/from step edges.\cite{Stout} The energy barrier for a particular atom to make an activated Monte Carlo move was taken to be proportional to the product of the coordination numbers of  the initial state site and the final state site. For growth onto surfaces vicinal to 6H SiC\{$000\overline{1}$\}, Stout's simulations bunched an initial train of single bilayer steps into a train where two nearby single bilayer steps accompany a 4 bilayer height step. No bunching into 3 bilayer height steps or 6 bilayer height steps was observed.

Finally, Camarda and co-workers used a full lattice KMC model including  defect sites to study step flow growth onto surfaces vicinal to 4H SiC(0001).\cite{Camarda} These authors did not treat  silicon and carbon atoms as diffusing species; the smallest growth unit considered was a Si-C dimer. Bunching was observed, but the step heights were not reported.

\section{Simulation Model}
We have developed a three-dimensional KMC simulation model based on the crystal structure of SiC. In this paper, the model is used to study step-flow etching and  step-flow growth of surfaces vicinal to 6H-SiC\{0001\}.
%6H SiC(0001) and 6H SiC($000\overline{1}$).
Later work will address island nucleation and multilayer roughness on singular surfaces and thermal decomposition of stepped and flat surfaces to produce epitaxial graphene. The starting vicinal surface studied was usually a uniform train of  36 single bilayer steps with a miscut angle of $\sim 15^\circ$($25^\circ$) for miscut oriented towards $<11 \overline{2} 0>$($<1 \overline{1} 00>$) . Otherwise (see Figure~\ref{fig:directionsandunits}), miscuts oriented towards $[1 \overline{1} 00]$ were treated using  ``helicoidal'' boundary conditions (HBC)\cite{Camarda2} along the $[10 \overline{1} 0]$ direction and periodic boundary conditions (PBC) along the $[11 \overline{2} 0]$ direction. The typical system size was $216 L_{1} \times 40 L_{2}$. Miscuts oriented towards $[\overline{1} 2 \overline{1} 0]$ were treated using  HBC along the $[11 \overline{2} 0]$ direction and PBC along the $[10 \overline{1} 0]$ direction. The typical system size was $72 L_{1} \times 216 L_{2}$.

The standard KMC method\cite{BKL}  identifies a set of  elementary ``moves'' and catalogs their relative rates. For  our simulations, thermal desorption directly into the gas phase was not allowed, but all atoms (except fully coordinated bulk atoms) were  permitted to move to empty nearest-neighbor or next-nearest-neighbor surface sites (with equal probability) at a rate $R=R_0\exp(-E/k_BT)$, where $R_0 = 10^{13}/{\rm sec}$, $k_B$ is Boltzmann's constant, and $T$ is the substrate temperature ( $T\simeq1000$~K for most of our simulations). The activation energy $E$  depends on the atom type and its local coordination through a bond-counting rule that includes only the four possible nearest neighbors. All the simulations we report used
\begin{eqnarray}
E_{Si} &=& \sum_{\rm 1st~ coord. sphere}E_{\rm Si-C} + \sum_{\rm 1st~ coord. sphere}E_{\rm Si-Si} \nonumber \\ \\
\nonumber \\\label{1stcoordsphbar}
E_C &=& \sum_{\rm 1st~ coord. sphere}E_{\rm Si-C} + \sum_{\rm 1st~ coord. sphere}E_{\rm C-C},\nonumber \\ \nonumber
\end{eqnarray}
with $E_{\rm Si-C} =$ 0.75~eV, $E_{\rm Si-Si}=$ 0.35~eV, and $E_{\rm C-C} = $ 0.65~eV. The absolute values of these parameters are not crucial because they only represent {\it  effective} pair-bond energies.  What matters is their relative ordering, which reflects (i) the stability of the SiC crystal and (ii) the much greater strength of the C-C bond compared to the Si-Si bond.
\begin{figure}
\begin{center}\includegraphics[
width=8.4cm, keepaspectratio] {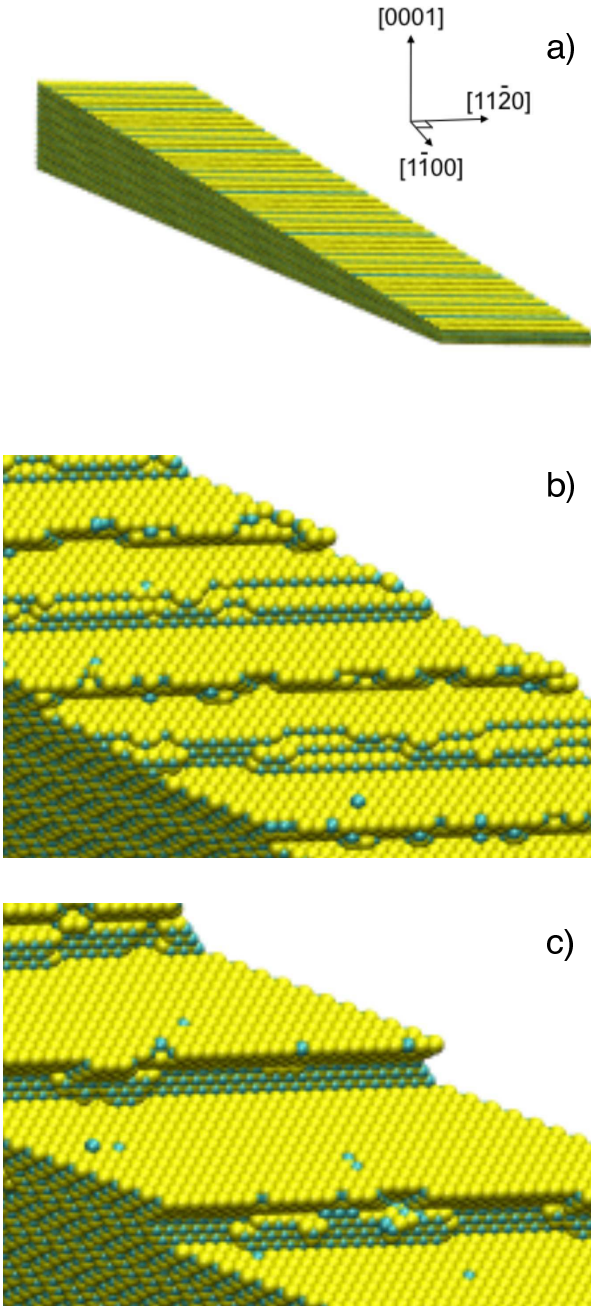}
\caption{\label{fig:growth[1m100]}  {Growth-induced step bunching on vicinal 6H-SiC(0001) surface with miscut towards $[1 \overline{1} 00]$.}}
\end{center}
\end{figure}

\begin{figure}
\begin{center}\includegraphics[
width=8.4cm, keepaspectratio] {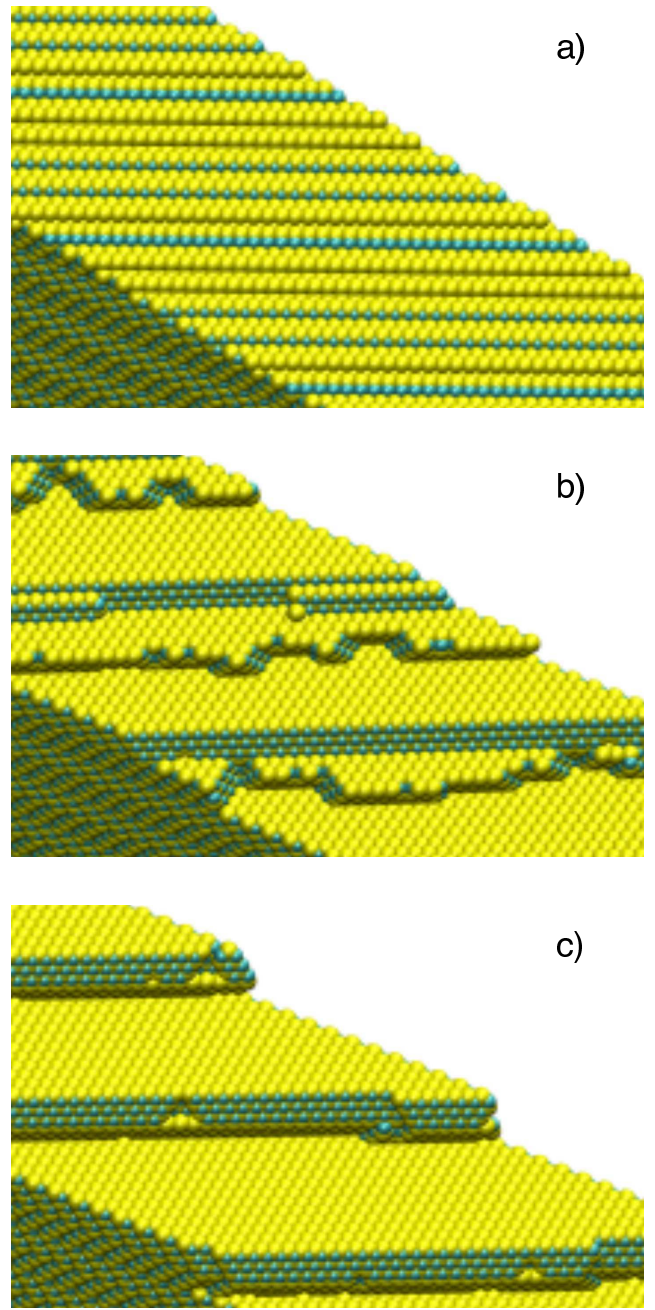}
\caption{\label{fig:etching[1m100]}  {Etching-induced step bunching on vicinal 6H-SiC(0001) surface with miscut towards $[1 \overline{1} 00]$.}}
\end{center}
\end{figure}

We come now to the crucial feature that distinguishes our simulations from others. For one species type (C or Si),  the surface jump rate computed using Eq.~(1) is exactly the same when the atom sits on any of the six 6H $(0001)$-type terraces (called $A$, $B$, $C$, $A^\ast$, $C^\ast$, and $B^\ast$ in Figure~\ref{fig:6H-SiCatomicarrangement}) exposed by a  vicinal surface with only single bilayer steps. However, the beyond nearest-neighbor interactions that energetically distinguish the many different polytypes of  SiC from one other imply that the energy barriers to surface migration cannot be exactly the same for all six terraces. The relevant surface diffusion barriers have not been reliably computed or measured. However, the first-principles, surface total energy calculations of Righi et al.\cite{Righi} show a clear energetic preference for SiC(0001)  surfaces to continue their subsurface stacking order. This conclusion agrees  with the observation that 3C polytype islands nucleate on 4H- and 6H-SiC(0001) substrates and with observed stable surface terminations for these exposed faces.\cite{Starke, Heinz, Nakagawa} Accordingly, we use the surface energy ordering computed in Ref.~\onlinecite{Righi} to scale the energy barriers for terrace surface diffusion. The scaling factors used were 1.0 for the atoms sitting on $A$ and $A^\ast$ terraces, 1.15 for the atoms sitting on $C$ and $B^\ast$ terraces, and 1.3 for the atoms sitting on $B$ and $C^\ast$ terraces. This procedure is consistent with our use of the binding energies in Eq.~(1) to estimate the bare energy barrier. Our choice of the scaling factors, or, more exactly, their ordering, is in agreement with the relative stability of the inequivalent terraces implied by the three different step velocities observed during the step flow growth of graphene by the decomposition of vicinal 6H-SiC(0001).\cite{Hupalo}

Etching was simulated very simply.  Every time an atom was selected to move along the surface, it was instead removed entirely from the simulation with a probability $\epsilon$, where  $0<\varepsilon \le 1$. The etching results were quite insensitive to the exact value of $\epsilon$. To study growth, we ignored precursor effects and deposited silicon and carbon with equal probability at randomly chosen empty surface sites of the SiC lattice. The average deposition rate used, $F=100/{\rm s}$, corresponds to a typical SiC growth rate of $\sim$ 1~$\mu$m/h.

\section {Results}
Our model produces very similar results for growth and etching of the Si-terminated face and C-terminated face of 6H SiC. This differs from the experimental observations summarized in Section II. For that reason, this section reports results only for the Si-terminated face.  The C-terminated face is discussed in Section V(D).
\subsection {Si-face with vicinal miscut towards $<1 \overline{1} 00>$}
Figure~\ref{fig:growth[1m100]} shows a sequence of simulated morphologies during step flow {\it growth} onto a vicinal 6H-SiC(0001) surface with the miscut towards the $[1 \overline{1} 00]$ direction. The step bunches that eventually form have the height of one  6H unit cell (6 bilayers). Note, however, that $\frac{1}{2}$ unit cell height steps (3 bilayers) form first. The faster growing bunch ($A^{\ast} C^{\ast} B^{\ast}$) catches up with the slower growing bunch (ABC) to form the final full unit cell height bunch. The step-edges of the final bunches are mostly straight and aligned along $[11 \overline{2} 0]$ direction, which is the energetically most stable configuration. The overhanging step risers produced by the simulation is an artifact of the nearest-neighbor approximation used in (1). A less steep (and smoother) step occurs when next-nearest-neighbors are included, but at the cost of a much slower simulation. This change introduces no qualitative effects on the bunching, so we used only the simpler model in this paper.

\begin{figure}
\begin{center}\includegraphics[
width=8.4cm, keepaspectratio] {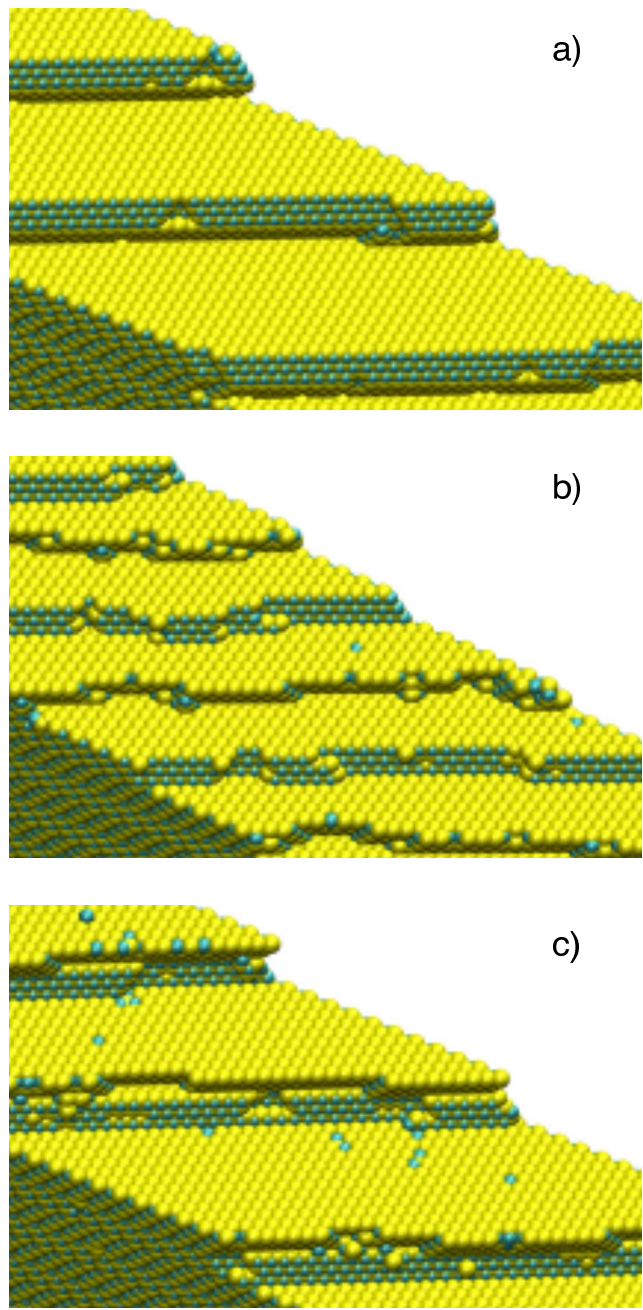}
\caption{\label{fig:growthafteretching[1m100]}  {Growth-induced step bunching on vicinal 6H-SiC(0001) surface with miscut towards $[1 \overline{1} 00]$. Initial surface configuration corresponds to etching-induced train of one unit cell height steps.}}
\end{center}
\end{figure}

For comparison with Figure~\ref{fig:growth[1m100]}, Figure~\ref{fig:etching[1m100]} shows a sequence of simulated morphologies during the step flow {\it etching} of a vicinal 6H-SiC(0001) surface with the miscut towards $[1 \overline{1} 00]$ direction.  The etching morphology we obtain is similar to the growth morphology except that the  ($A^{\ast} C^{\ast} B^{\ast}$) bunch retracts faster than (ABC), and ends up at the bottom of the full unit cell height step. During growth, the ($A^{\ast} C^{\ast} B^{\ast}$) bunch winds up on top of the (ABC) bunch.
It is interesting to note that many experiments precede epitaxial growth of 6H-SiC with a gas etching step to smoothen the surface. To study this case, we performed a growth simulation beginning with the  surface shown in the last panel of Figure~\ref{fig:etching[1m100]}.
As shown in Figure~\ref{fig:growthafteretching[1m100]},  the starting full unit  cell height step with (ABC) on top of $(A^\ast C^\ast B^\ast$) flips during the growth to a full unit cell height step with  $(A^\ast C^\ast B^\ast$) on top of (ABC). In other words, the starting surface does not matter and the  final panel of Figure~\ref{fig:growth[1m100]} is the same as the final panel of Figure~\ref{fig:growthafteretching[1m100]}.

\begin{figure}
\begin{center}\includegraphics[
width=8.4cm, keepaspectratio] {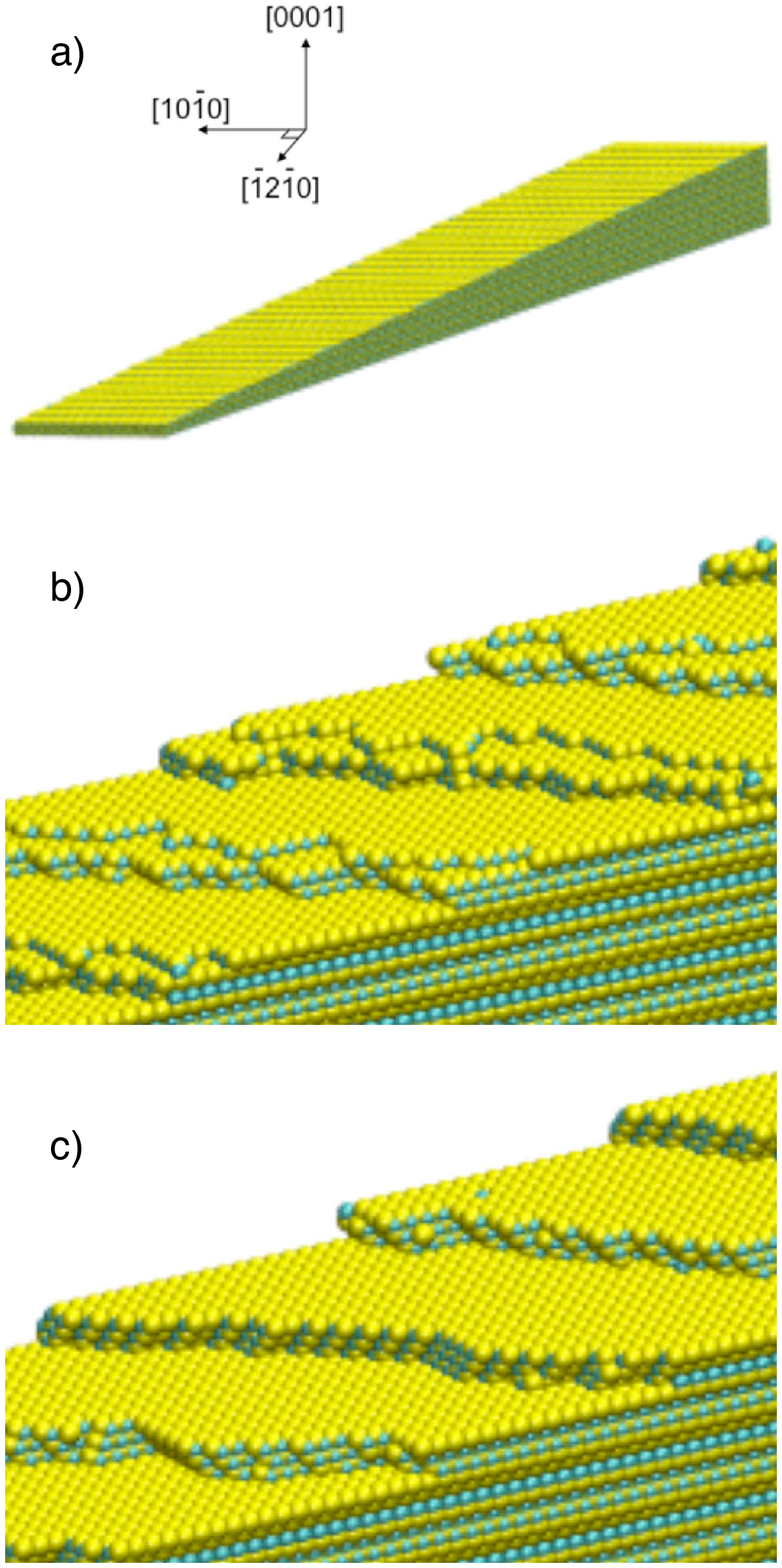}
\caption{\label{fig:growth[11m20]}  {Growth-induced step bunching on vicinal 6H-SiC(0001) surface with miscut towards $[11 \overline{2} 0]$.}}
\end{center}
\end{figure}

\begin{figure}
\begin{center}\includegraphics[
width=8.4cm, keepaspectratio] {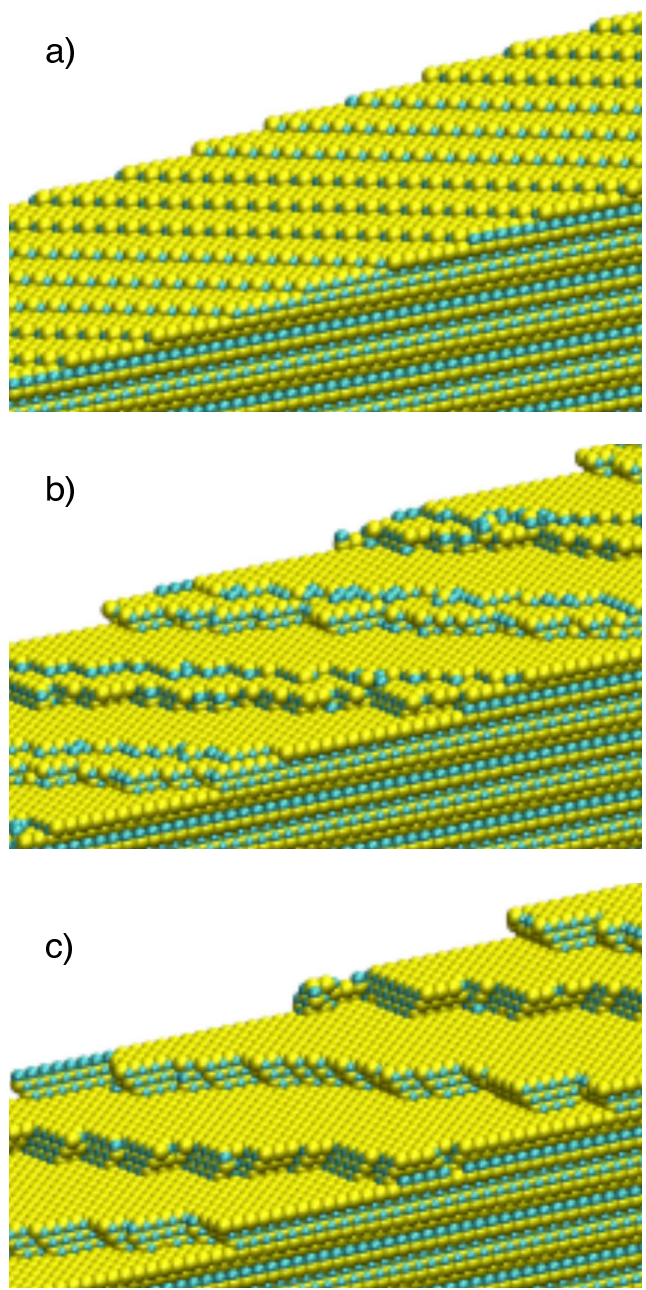}
\caption{\label{fig:etching[11m20]}  {Etching-induced step bunching on vicinal 6H-SiC(0001) surface with miscut towards $[11 \overline{2} 0]$.}}
\end{center}
\end{figure}

\subsection {Si-face with vicinal miscut towards  $<11 \overline{2} 0>$ }
Figure~\ref{fig:growth[11m20]} shows a sequence of simulated morphologies during  step flow growth  onto a vicinal 6H-SiC(0001) surface with the miscut towards the
$[\overline{1} 2 \overline{1} 0]$ direction. We observe the formation of $\frac{1}{2}$ unit cell height steps with zigzag shaped step edges.  On average, the steps are aligned along the $[10 \overline{1} 0]$ direction (perpendicular to the miscut direction). However, the straight segments of the step edges are aligned along the close-packed $<11 \overline{2} 0>$ directions. Etching produces very similar results in the sense that $\frac{1}{2}$ unit cell height steps form. This is shown in Figure~\ref{fig:etching[11m20]}.

\section {Discussion}
\begin{figure}
\begin{center}\includegraphics[
width=8.4cm, keepaspectratio] {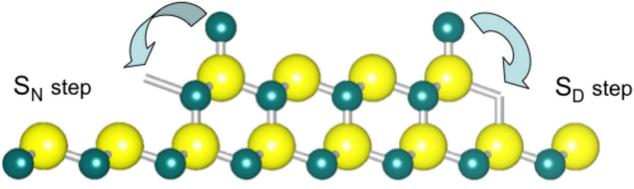}
\caption{\label{fig:SNandSDsteps}  {Side view of $S_{N}$ and $S_{D}$ steps. See text for discussion.}}
\end{center}
\end{figure}

The results of our KMC simulations are in good qualitative agreement with experiments. The formation of full unit cell height steps (six bilayers) is inherent in SiC step-flow growth/etching on vicinal 6H-SiC(0001) surfaces with the miscut towards $<1 \overline{1} 00>$. On the other hand, $\frac{1}{2}$ unit cell height steps form during simulations of growth/etching when the miscut is along $<11\overline{2}0>$. We understand all these features in terms of the six different steps that appear on any surface composed of only bilayer height steps (top panel of Figure~\ref{fig:etching[1m100]} or Figure~\ref{fig:etching[11m20]}). There are six different velocities because there are three inequivalent terraces where surface diffusion events occur and two inequivalent step edges, where attachment, detachment, and interlayer transport events occur. As described in Section~III, the surface diffusion rate is fastest for adatoms sitting on A and A$^\ast$ terraces, slower for adatoms sitting on C and B$^\ast$ terraces, and slowest for adatoms sitting on B and C$^\ast$ terraces.
Figure~\ref{fig:SNandSDsteps} shows the two types of steps.\cite{Pechman, Ramachandran} For the $S_N$ step, a next-to-next nearest neighbor jump is required for an atom to attach to the step from its upper bounding terrace. Moreover, after this jump occurs, the attached atom is only singly bonded to the step, and thus easily detached. For the  $S_D$ step, only a  next-nearest neighbor jump is required for an atom to attach to the step from its upper bounding terrace. After this jump occurs, the attached atom is doubly  bonded to the step, and thus less likely to detach.

\begin{figure}
\begin{center}\includegraphics[
width=8.4cm, keepaspectratio] {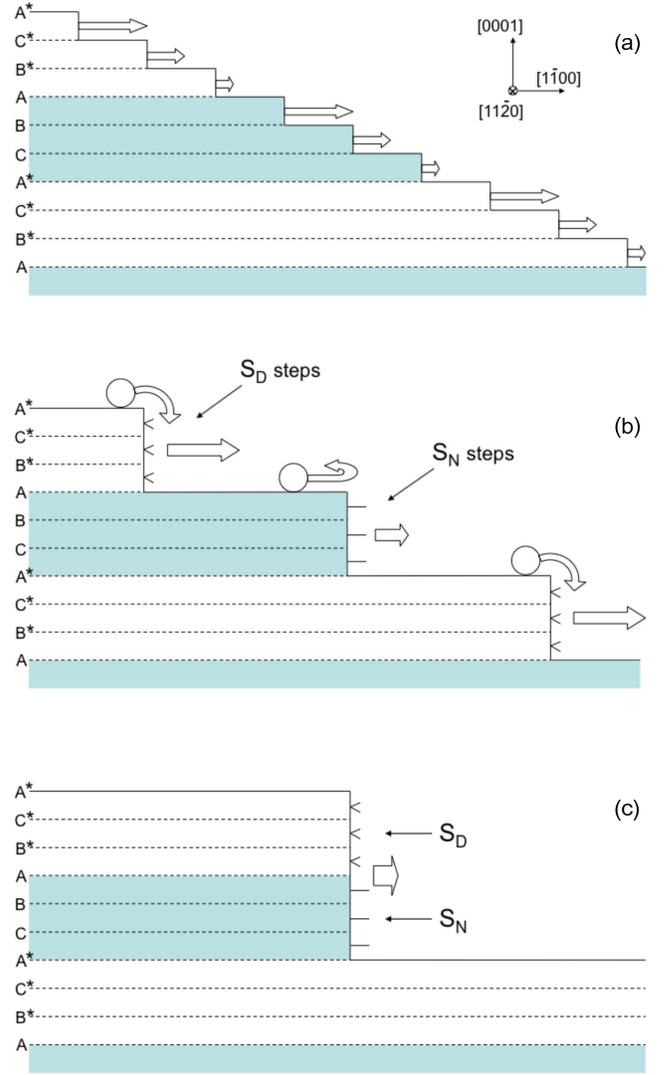}
\caption{\label{fig:stepflowgrowth}  {A cartoon of step-flow growth on 6H SiC(0001): (a) Different length arrows indicate the different growth velocities of the  steps which terminate the three inequivalent bilayer terraces;  (b) Two types of $\frac{1}{2}$ unit cell height steps ($S_{N}$ and $S_{D}$) differ by the number of dangling bonds for the outermost step edge atoms. The presence  of an Ehrlich-Schwoebel energy barrier to downward interlayer diffusion at $S_{N}$ steps explains the difference in growth speed between $S_{N}$ and $S_{D}$ $\frac{1}{2}$ unit cell height step bunches; (c)  $S_{D}$ steps wind up on top $S_{N}$ steps to form a single, full unit cell height step.}}
\end{center}
\end{figure}

Figure~\ref{fig:stepflowgrowth} and Figure~\ref{fig:stepflowetching} show scenarios for growth and etching, respectively. The fastest steps during growth are bounded from above by the most stable terraces (A and A$^\ast$) . The slowest steps during growth  are bounded from above by the least stable terraces (C and B$^\ast$). For the same reason of stability,  fast growing steps are the slowest etching and vice versa. Accordingly, the terraces exposed at the time when $\frac{1}{2}$ unit cell height steps are present on the surface, in both cases correspond to A and $A^{\ast}$ (see Figure~\ref{fig:6H-SiCatomicarrangement}), which is in agreement with experiments.\cite{Hayashi}

\subsection{Step-Flow Growth}
The top panel of Figure~\ref{fig:stepflowgrowth} shows a vicinal surface composed of single bilayer steps. The arrows on the steps reflect their relative velocities due to the rates of surface diffusion on the three inequivalent terraces mentioned just above. As a result, the single bilayer steps bunch into the ${1\over 2}$ unit cell height steps shown in the second panel of Figure~\ref{fig:stepflowgrowth}. The subsequent bunching of these steps into the full unit cell height step shown in the final panel of Figure~\ref{fig:stepflowgrowth} occurs because one half cell height step is  $S_N$ type and the other is  $S_D$ type. The latter moves faster than the former because the next-to-next nearest neighbor jump required for attachment to the $S_N$ step from above is not allowed in our model. In reality, we presume there is simply a higher barrier for this process to occur at a $S_N$ step than at a $S_D$ step. In other words, the Ehrlich-Schwoebel barriers associated with these steps are different.\cite{Ehrlich, Michely}

\begin{figure}
\includegraphics[
width=8.4cm, keepaspectratio] {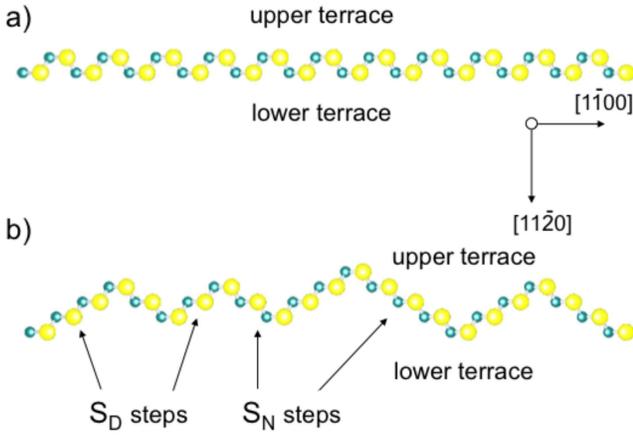}
\caption{\label{fig:1120steptransform}  {(a) A step edge of the so called ``open" step (top view), aligned along $<1 \overline{1} 00>$, is shown schematically. (b) Growth and etching at such steps typically results in development of triangular protrusions. The alternating straight segments of this protrusions ($S_{N}$ and $S_{D}$ step edges) are aligned along the $<11 \overline{2} 0>$ directions.}}
\end{figure}

The preceding discussion applies directly for vicinal miscuts toward the $<1\bar{1}00>$ direction. However, we have stated that full unit cell height steps do not form in our SiC growth simulations for miscuts toward the $<11 \overline{2} 0>$ direction. This occurs because
the step edges in this case have a natural zigzag shape, consisting of alternating straight segments, corresponding to $S_{N}$, or $S_{D}$ steps (see Figure~\ref{fig:1120steptransform}). Each step edge has equal portions of $S_{N}$ and $S_{D}$ steps, and, as a result, all $\frac{1}{2}$ unit cell height steps propagate with the same speed. Of course, if the miscut is not exactly toward the $<11 \overline{2} 0>$ direction, adjacent $\frac{1}{2}$ unit cell height steps differ in their relative population of $S_{N}$ and $S_{D}$ step edges. This may trigger the formation of full unit cell height steps.

\subsection{Step Flow Etching}
\begin{figure}
\begin{center}\includegraphics[
width=8.4cm, keepaspectratio] {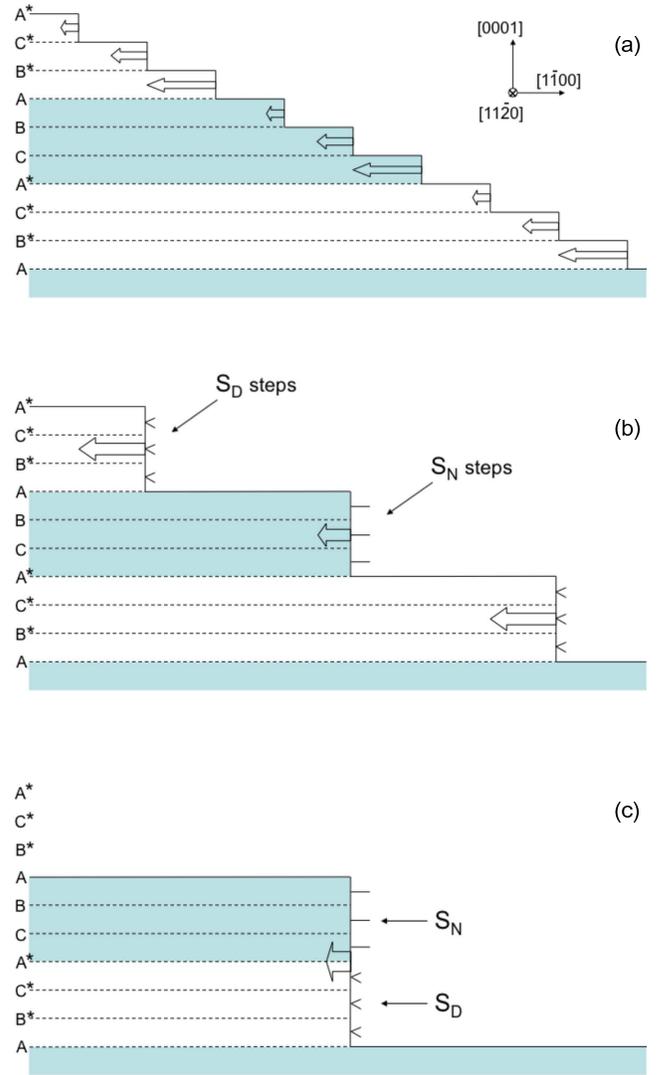}
\caption{\label{fig:stepflowetching}  {A cartoon of step-flow etching of 6H SiC(0001): (a) Different length arrows indicate the different etch velocities of the  steps which terminate the three inequivalent bilayer terraces;  (b) Two types of $\frac{1}{2}$ unit cell height steps ($S_{N}$ and $S_{D}$) have different edge velocities because they differ in the number of dangling bonds for the outermost step edge atoms; (c)  $S_{N}$ steps wind up on top $S_{D}$ steps to form a single, full unit cell height step.}}
\end{center}
\end{figure}
Etching is often regarded as the inverse of growth. \cite{Nakamura, Feenstra}
Therefore, mimicking our discussion of growth, the top panel of Figure~\ref{fig:stepflowetching} shows a vicinal surface composed of single bilayer steps. The arrows on the steps reflect their etch velocities due to the different step detachment rates associated with the three inequivalent terraces discussed earlier. As a result, the steps  bunch to form the ${1\over 2}$ unit cell height steps shown in the second panel of Figure~\ref{fig:stepflowetching}. The subsequent bunching of these steps into the full unit cell height step shown in the final panel of Figure~\ref{fig:stepflowgrowth} occurs because the $S_N$-type step, which has triply bonded outermost atoms, etches more slowly than the $S_D$-type step, which has only doubly bonded outermost atoms.

\begin{figure}
\includegraphics[
width=8.4cm, keepaspectratio] {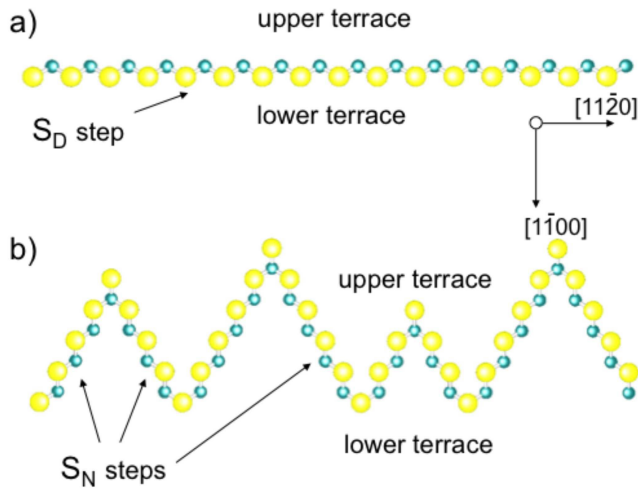}
\caption{\label{fig:SDsteptransform}  {(a) $S_{D}$ step edge (top view) is shown schematically. (b) $S_{D}$ steps are less stable then $S_{N}$ steps, because the outermost atoms of the former have two nearest-neighbor bonds, while the outermost atoms of the latter have three such bonds. Growth and etching is faster at $S_{D}$ steps and typically results in development of triangular protrusions. The straight segments of this protrusions are aligned along the $<11 \overline{2} 0>$ directions and correspond to energetically stable $S_{N}$ steps.}}
\end{figure}

In agreement with experiments, we observe the development of ``triangular" protrusions (see Figure~\ref{fig:etching[1m100]}) which form as a result of etching at $S_{D}$ step bunches.\cite{Ramachandran}  As Figure~\ref{fig:SDsteptransform} shows, the straight segments of these protrusions are aligned at angles of $30^{\circ}$ with respect to the direction of miscut ($[1 \overline{1} 00]$) and thus correspond to energetically stable $S_{N}$ step bunches. As discussed in detail in Ref.~\onlinecite{Ramachandran}, the outermost atoms of these protrusions, which have only two bonds with nearest neighbors, constitute another source of instability. As a result, preferential etching of these protrusions leads to the formation of full unit cell height steps.

In our simulations of SiC etching for miscut towards $<11 \overline{2} 0>$, we do not observe the formation of unit cell height steps. This agrees with the most recent experimental results. \cite{Nakajima, Feenstra, Hayashi} We explain this in terms of the previously discussed zigzag structure of the steps which occur on this surface.  Each $\frac{1}{2}$ unit cell height step with a zigzag shape has equal portions of faster and slower etched straight segments, which correspond to $S_{D}$ and $S_{N}$ step bunches. For this reason the etching rates of all $\frac{1}{2}$ unit cell height step are identical (on average) and unit cell height steps do not form.

Notwithstanding the foregoing, some etching experiments on surfaces with
miscut towards $<11 \overline{2} 0>$ {\it do} see the formation of full unit cell height steps. \cite{Nakamura, Nakagawa, Hayashi}  A possible explanation for these conflicting observations is the previously mentioned possibility of deviations of the miscut from exactly   $<11 \overline{2} 0>$ with its attendant steps with faster and slower etching segments. When the populations of these segments is not equal, the  $\frac{1}{2}$ unit cell height steps with more fast-etching segments catch up to the steps with fewer fast-etching segments.\cite{Nakajima, Hayashi} The result is a train of full  unit cell height steps.

\subsection{Surface Terminations}
Only one type of terrace is exposed to the vacuum after all full unit cell steps have formed  on a $[1 \overline{1} 00]$ miscut surface. However, our KMC simulations predict that the exposed terrace is not the same after growth as it is after etching.  After growth, the $A^{\ast} C^{\ast} B^{\ast}$-bunch (the outermost atoms of bilayer-steps have two bonds with nearest neighbors and two dangling bonds) is on top of the ABC-bunch (the outermost atoms of bilayer-steps have three bonds with nearest neighbors and one dangling bond). This is called the $S_{3}^{*}$  surface termination in the literature.\cite{Hayashi} After etching, the sequence of bilayers at the surface is
opposite: $... B^{\ast}C^{\ast}A^{\ast}CBA$.
This is called the $S_{3}$ surface termination. Cross section TEM experiments could be used to test this prediction.
\subsection{The C-terminated surface}
The simulation results we have presented so far describe the evolution of surface morphology during the epitaxial growth/etching of 6H-SiC on the vicinal 6H-SiC(0001) surface (Si-terminated face). Our simulation results for the C-terminated face (6H-SiC(000$\overline{1}$)), using the same model parameters, are qualitatively very similar. This disagrees with the step bunching behavior  observed in experiments which is typically less pronounced for the C-face compared to the Si-face.\cite{Kimoto, Feenstra} On the other hand, the C-face data are quite explicable if the terrace diffusion rates and step mobilities  differ from their values on the Si-face.

The terrace diffusion scaling factors used to take account of the three inequivalent terraces of SiC(0001) were  chosen based on the Si-face calculations of Righi {\it et. al}.\cite{Righi} These  authors  did not perform similar calculations for the C-face and it is  possible that the results are different. One possibility is that the scale factor ordering is the same as for the Si-face, but that their magnitudes are less different. Another possibility is that the ordering of the scale factors differs on the C-face. Experimental support for this comes from the different surface termination observed for the  two polar faces of 6H-SiC\{0001\}.  The so-called $(2\times2)_{C}$ reconstruction, which stabilizes hexagonal stacking at the surface ( $S_{1}$ stable surface termination) is sometimes observed on the C-face.\cite{Seubert} This implies a different ordering for the three inequivalent terrace scaling factors than for the Si-face.

It seems quite likely that the Ehrlich-Schwoebel barriers to interlayer diffusion (which strongly influence the bunching of half unit cell height steps into full unit cell height steps in our model) are different on the C-face and the Si-face. The magnitude of these barriers is intimately connected to the structure of the steps and there is theoretical evidence that the step structure indeed depends on the polarity of the surface.\cite{Pearson}   If the difference between the corresponding barriers is less pronounced for the C-face, we would find the experimentally observed delay in the formation of full unit cell height steps during the growth of 6H-SiC on the vicinal 6H-SiC(000$\overline{1}$) surface. Moreover, strong barriers to interlayer diffusion at both $S_{N}$ and $S_{D}$ steps suppress interlayer mass transport and stabilize the persistence of single bilayer steps.  This would explain the observed experimentally preference of bilayer height steps on C-terminated surfaces to remain completely unbunched during growth.\cite{Kimoto}

Finally, it is possible that step bunching is less pronounced on the C-face because the elastic-driven repulsive interaction between the steps of a vicinal surface\cite{Muller} is more  pronounced on the C-face compared to the Si-face. The repulsion depends on the step stiffness,\cite{Marchenko} which in turn depends on the step structure, which are doubtless different for the two faces.

\section {Conclusions}
We have used lattice KMC simulations to study the formation of step bunches during growth and etching of 6H-SiC(0001) vicinal surfaces. For both situations, the simulations show that single bilayer steps bunch into half unit cell steps (3 bilayers each), which subsequently bunch into full unit cell steps. This is consistent with experimental observations for both the Si-terminated face and the C-terminated face except that we obtain greater bunching for the C-face than seen in experiment. The main driving force for bunching into half unit cell height steps is that surface diffusion is not equally fast on all bilayer terraces. The main driving force for the subsequent bunching into full height unit cells is the existence of two different local atomic step structures, which leads to two different step mobilities. A prediction of the model which invites an experimental test is that growth-induced and etching-induced step bunching lead to different surface terminations for the exposed terraces when full unit cell steps are present.

\begin{acknowledgments}
The authors acknowledge helpful correspondence with Randall Feenstra and Miron Hupalo. The work of V.B. was supported by the Department of Energy under Grant No. DE-FG02-04-ER46170. We also acknowledge a grant of computer time from the National Center for Supercomputing Applications.
\end{acknowledgments}

\end{document}